\documentclass[prb,aps,onecolumn,,superscriptaddress]{revtex4-1}
 \newcommand{\beq}{\begin{equation}}
\newcommand{\eeq}{\end{equation}}
\usepackage{graphics}
\usepackage{epsf}
\usepackage{amsmath}
\usepackage{amsfonts}
\usepackage{amssymb}
\usepackage{graphicx}
\usepackage{epsfig}
\usepackage{color}

\begin{document}

\title{ Epitaxial stabilization of ultra thin films of electron doped manganites}
\author {S. Middey}
\email{smiddey@uark.edu  }
\affiliation  {Department of Physics, University of Arkansas, Fayetteville, Arkansas 72701, USA}
\author{M. Kareev}
\affiliation  {Department of Physics, University of Arkansas, Fayetteville, Arkansas 72701, USA}
\author{D. Meyers}
\affiliation  {Department of Physics, University of Arkansas, Fayetteville, Arkansas 72701, USA}
\author{X. Liu}
\affiliation  {Department of Physics, University of Arkansas, Fayetteville, Arkansas 72701, USA}
\author{Y. Cao}
\affiliation  {Department of Physics, University of Arkansas, Fayetteville, Arkansas 72701, USA}
\author{S. Tripathi}
\affiliation  {Department of Physics, University of Arkansas, Fayetteville, Arkansas 72701, USA}
\author{P. Ryan}
\affiliation {Advanced Photon Source, Argonne National Laboratory, Argonne, Illinois 60439, USA}
\author{J. W. Freeland}
\affiliation {Advanced Photon Source, Argonne National Laboratory, Argonne, Illinois 60439, USA}
\author{ J. Chakhalian}
\affiliation  {Department of Physics, University of Arkansas, Fayetteville, Arkansas 72701, USA}

\begin{abstract}
Ultra-thin films of the electron doped manganite La$_{0.8}$Ce$_{0.2}$MnO$_3$ were grown in a layer-by-layer growth mode on SrTiO$_3$ (001) substrates by pulsed laser interval deposition. High structural quality  and surface morphology was confirmed by a combination of synchrotron based x-ray diffraction and atomic force microscopy. Resonant X-ray absorption spectroscopy measurements confirm the presence of Ce$^{+4}$ and Mn$^{+2}$ ions. In addition,  the  electron doping signature was corroborated by Hall effect measurements. All grown films show ferromagnetic ground state  as revealed by  both XMCD and magnetoresistance measurements and remain insulating contrary to  earlier reports of metal-insulator transition. Our results hint at  the possibility of electron-hole asymmetry in the colossal magnetoresistive (CMR) manganite phase diagram  akin to  high-$T_c$ cuprates.

\end{abstract}

\maketitle

Hole-doped manganites ($R_{1-x}A_x$MnO$_3$) have been extensively studied  over the last few decades~\cite{Dagotto,Salamon,tokura1} not only for the realization of different exotic phenomena ($e.g.$ colossal  magnetoresistance, phase separation, charge ordering  etc.) driven by the strong coupling among the lattice, spin, charge, and orbitals degrees of freedom but also for the huge technological potential of using these materials in spintronic devices~\cite{devices,rmp_devices,Yajima}. While the hole doping ($i.e.$ increase in Mn oxidation state) can be realized by the partial substitution of $R^{+3}$ ions of $R$Mn$^{+3}$O$_3$ by some $A^{+2}$ ions, the successful replacement of $R^{+3}$ by some $A^{+4}$  cations should reduce the same amount of Mn to +2 charge state.  The  electron doping may result in  an asymmetric phase diagram, as for instance has been established  for the case of high $T_c$ cuprate family~\cite{kotlier,cuprates,edopedcuprates}. 
However, unlike  the hole doping in cuprates, where the holes are known to reside on the ligand (oxygen)  $p$ orbitals~\cite{holesonligand}, the electrons can be directly doped into the Cu $d$ orbital  states~\cite{edopedxas} resulting  in a very distinct phase  diagram with doping concentration. Towards this end, though the synthesis of the electron doped manganites was   expected   to  be  similar to that of the electron-doped cuprates~\cite{edoped_tokura} using Ce as dopant,  the experimental reports~\cite{lcemo_bulk_mandal,lcemofilm,lcemofilm_phasediagram,lcemo_xas_edope,pndiode_apl,tmr_prl,Venkatesan_JMMM,cecluster,hall_diffoxygencontent,xmcd,transport} featuring Ce doped LaMnO$_3$, are  in sharp conflict with each other questioning the feasibility of reliable electron doping for the manganite compounds to  investigate this fundamental issue.

Specifically, after the first report of synthesizing Ce doped $R$MnO$_3$ ($R$ = La, Pr, Nd) in polycrystalline powder form~\cite{lcemo_bulk_mandal}, the phase diagram of La$_{1-x}$Ce$_x$MnO$_3$ thin film was studied as a function of $x$~\cite{lcemofilm_phasediagram}. It is interesting  to  note that, contrary to the highly asymmetric  $T-x$ phase diagram of high $T_c$ cuprates family~\cite{cuprates}, the phase diagram of manganites (LaR)MnO$_3$ as a function of hole~\cite{Cheong} and electron doping~\cite{lcemofilm_phasediagram}  appeared to be surprisingly symmetric. While the electron doping in La$_{0.7}$Ce$_{0.3}$MnO$_3$ was conjectured from the observation of Ce$^{+4}$ valency in the XAS spectrum~\cite{lcemo_xas_edope} and from  experiments with a  $p-n$ junction composed of   hole and electron doped manganites layers ~\cite{pndiode_apl,tmr_prl},  the latter reports~\cite{xmcd,transport} have argued that the ferromagnetism and insulator to metal transition (IMT)  may arise from uncontrolled $hole$ doping. In addition, the Hall effect measurements on La$_{0.7}$Ce$_3$MnO$_{3+\delta}$ with varying oxygen content $\delta$  demonstrated ~\cite{hall_diffoxygencontent} that the samples with electrons as the carrier type lack any IMT and  remain insulating throughout the whole temperature range.  Those  contradicting reports have intensified the overall concern about the true nature of electronic  and magnetic ground state in electron doped manganites generated by the Ce doping.

In this letter, we have report on the growth of high  quality thin films of La$_{0.8}$Ce$_{0.2}$MnO$_3$ on a SrTiO$_3$ (001) substrate using  pulsed laser interval deposition technique. While the extensive structural  and surface morphology characterization using reflection high energy electron diffraction (RHEED), atomic force microscopy (AFM), synchrotron based X-ray diffraction (XRD) confirmed excellent structural quality of the films; the presence of  electron doping was confirmed by resonant X-ray absorption spectroscopy on Mn $L_{3,2}$ and Ce $M_{4,5}$-edge and by Hall effect measurements. The presence of ferromagnetism was assessed by XMCD on Mn L-edge and by magnetoresistance measurements  up  to  7T. Contrary to the earlier reports of metal-insulator transition for this  composition, our electron doped samples remain insulating throughout the whole temperature range of measurement with ferromagnetic ground state. 
  
 
 La$_{0.8}$Ce$_{0.2}$MnO$_3$ films with  20 uc (unit cell) thickness were grown on high-quality SrTiO$_3$ (STO) (001) substrates (Crystec, Germany) by pulsed laser interval deposition~\cite{misha_jap} using a KrF excimer laser operating at $\lambda$ = 248 nm and 18 Hz pulse rate. The growth was carried out at 870$^\circ$C under  250 mTorr  partial pressure of oxygen  and the entire layer by layer growth process was monitored in-situ by  high pressure RHEED. In order to avoid the over oxidation of the samples by annealing at the growth temperature, the sample was first cooled down to 600$^\circ$C at a rate of 15$^\circ$C/min immediately after finishing the growth and then annealed  for 30 minutes under 500 Torr  of ultra pure oxygen. The X-ray diffraction patterns were recorded at the 6-ID-B beam line of the Advanced Photon Source (APS). In order to determine charge state of Mn and Ce,  resonant X-ray absorption spectra of Mn $L_{3,2}$ edge and Ce $M_{4,5}$ edges have been acquired at the 4-ID-C beam line of the APS in total electron yield (TEY) mode. The  thickness ($\sim$ 7.8 nm) of these films compared to the probing depth
($\sim$ 10 nm) of the TEY mode  allow us to probe the electronic and magnetic structure throughout the entire sample. The magneto-transport properties and Hall resistance in an applied magnetic  field up  to  7 T were measured in a Physical Property Measurement system (PPMS).

 The time dependent  intensity of the specular reflection of the RHEED pattern during  growth are shown in  Fig. 1(a). As seen, the intensity sharply drops during the rapid ablation and recovers within the next few seconds, confirming the layer-by-layer growth.~\cite{misha_jap,enogrowthpaper}  The characteristic specular (0,0) and off-specular reflection (0,1) \& (0,-1) Bragg reflections together with  the streaking patterns in the diffraction image (Fig. 1(b)) obtained after cooling to room temperature imply  smooth flat terraces.  The absence of any half-order peaks~\cite{enogrowthpaper} in the RHEED patterns implies that any orthorhombic distortion, if present, must be very small. The AFM image of the surface morphology of the film is shown in Fig. 1(c) yields an average surface roughness $\sim$ 95 pm, further affirming their high morphological quality. Additionally, the structural quality of the grown film has been investigated by  synchrotron based x-ray diffraction.  Fig. 1(d) displays the diffraction pattern obtained in the vicinity of the STO (0 0 1) reflection. As seen, the broad film peak, marked by the arrow corresponds to the out-of-plane lattice constant of  3.8663 \AA; the result is  consistent with the expected  tetragonal  distortion  under tensile strain for  the cube on cube growth. 

After confirming the high structural quality of the samples, we turn our attention to  the key  question of  electron doping. As Ce can have both +3 and +4 oxidation states, the key criterion of electron doping is the presence of  stabilized Ce$^{+4}$ in our film. To this end we used resonant  X-ray  absorption which is  a direct probe of  the charge oxidation state ~\cite{cexas}.   Fig. 2(a) shows the XAS pattern obtained at  the Ce $M_{4,5}$ edge along with with  the reference spectra  reported earlier from Ce(+4)O$_2$ and Ce(+3)F$_3$ standards~\cite{lcemo_xas_edope}. A direct comparison to  the Ce reference samples reveals  that  the LCeMO line shape is very similar to that  of  Ce(+4)O$_2$ standard. At the same time  the  LCeMO spectrum has  additional small features at the lower energy side for both $M_5$ and $M_4$ edges. Specifically, the additional low intensity shoulder around 886 eV  matches the  main peak of $M_5$ edge for  Ce(+3)F$_3$ implying the small contribution of Ce in the +3 oxidation state. This  implies that the magnitude of the electron doping concentration is  slightly lower  than nominal  20\% doping  expected from the formal chemical formulae. Complimentary  to  Ce XAS results, Fig. 2(b) shows the Mn $L_3$ edge XAS spectra for the LCeMO film and Mn(+2)O standard recorded simultaneously with the sample  in the diagnostic section of the beamline.  As shown, the shoulder position at 638.7 eV matches exactly with the position of the main peak of the MnO standard and thus clearly indicates the presence of $divalent$ Mn within the electron doped  LCeMO film. In short, the combined  presence of Ce$^{+4}$ and Mn$^{+2}$ charge states  directly establishes the formation of the electron doped phase of LCeMO.

With  the knowledge of  electronic  structure we precede with studies of magnetization and magneto-transport properties of the LCeMO. Towards this end, the  temperature-dependent resistivity behaviors, with and without external magnetic field were measured during cooling and heating cycles; since both curves are  identical only the cooling curves are shown in Fig. 3(a).  Contrary  to the earlier reports by Zhao $et. \ al.$~\cite{Venkatesan_JMMM} suggesting that La$_{0.8}$Ce$_{0.2}$MnO$_3$ films would  undergo a metal-insulator transition due to hole doping, the present film is insulating  at  all  temperatures and does not show any transition to the metallic state. To  exclude the effect of strain  we have grown 20 uc LCeMO film on YAlO$_{3}$ under large compressive strain. The resulting sample is still  insulating at  the whole  temperature range  (not shown) thus affirming  the robustness of the insulating ground state. The application of an external magnetic field results in large negative magnetoresistance (see Fig. 3(a) and inset), characteristic of colossal magnetoresistance.   In order to define the carrier type (hole vs. electron), the Hall effect has been measured by sweeping the magnetic field ($H$) between -7 T and +7 T at different temperatures. Figure 3(b) shows the dependence of transverse Hall resistivity ($\rho_{\textrm{xy}}$)~\cite{Hall} on $H$ . 
 As   observed, the negative sign of $R_H$  at 200 K  highlights the excellent agreement with our conclusion on  electron doping obtained from XAS measurements.  As a rough  estimate, assuming a Fermi surface with one type of charge carriers, the carrier concentration $n$ ($R_H$=1/$ne$) can be  estimated $\sim 3.6 \times 10^{20}$ electrons/cm$^3$. In passing we note, our Hall  resistance results are in variance with  the  earlier reports~\cite{hall_diffoxygencontent,transport} about positive Hall coefficient in Ce doped manganite.  In order to further confirm the presence of ferromagnetism, x-ray absorption spectra at the Mn $L_{3,2}$ edges were recorded with left and right circularly polarized light.  The strong XMCD (x-ray magnetic circular dichroism) signal (Fig. 3(c)) obtained at 150 K with an applied 5 Tesla magnetic field establishes the characteristic ferromagnetic response of Mn  in this electron doped sample.

In summary, we have developed  the layer-by-layer growth of high quality fully epitaxial films of electron doped LCeMO manganites. A combination of RHEED, synchrotron  based XRD and XAS confirmed the  excellent structural, chemical, and electronic quality, while transport measurements confirmed the insulating nature of the film.  The presence of ferromagnetism is determined from the large magnetoresistance and Mn XMCD signal.  The Hall effect measurements corroborates that electron as the charge carrier. The present study hints at  the possible electron-hole asymmetry in the CMR manganite phase diagram  and opens exciting possibilities of merging this electron doped material with the other hole doped systems to fabricate oxide based spintronic junctions.

J. C. was supported by DOD-ARO under Grant No. 0402-17291. Work at the Advanced Photon Source, Argonne was supported by the U.S. Department of Energy, Office of Science under Grant No. DEAC02-06CH11357.

 \newpage

  \begin{figure} [t!]
\includegraphics[width=3.5in,clip] {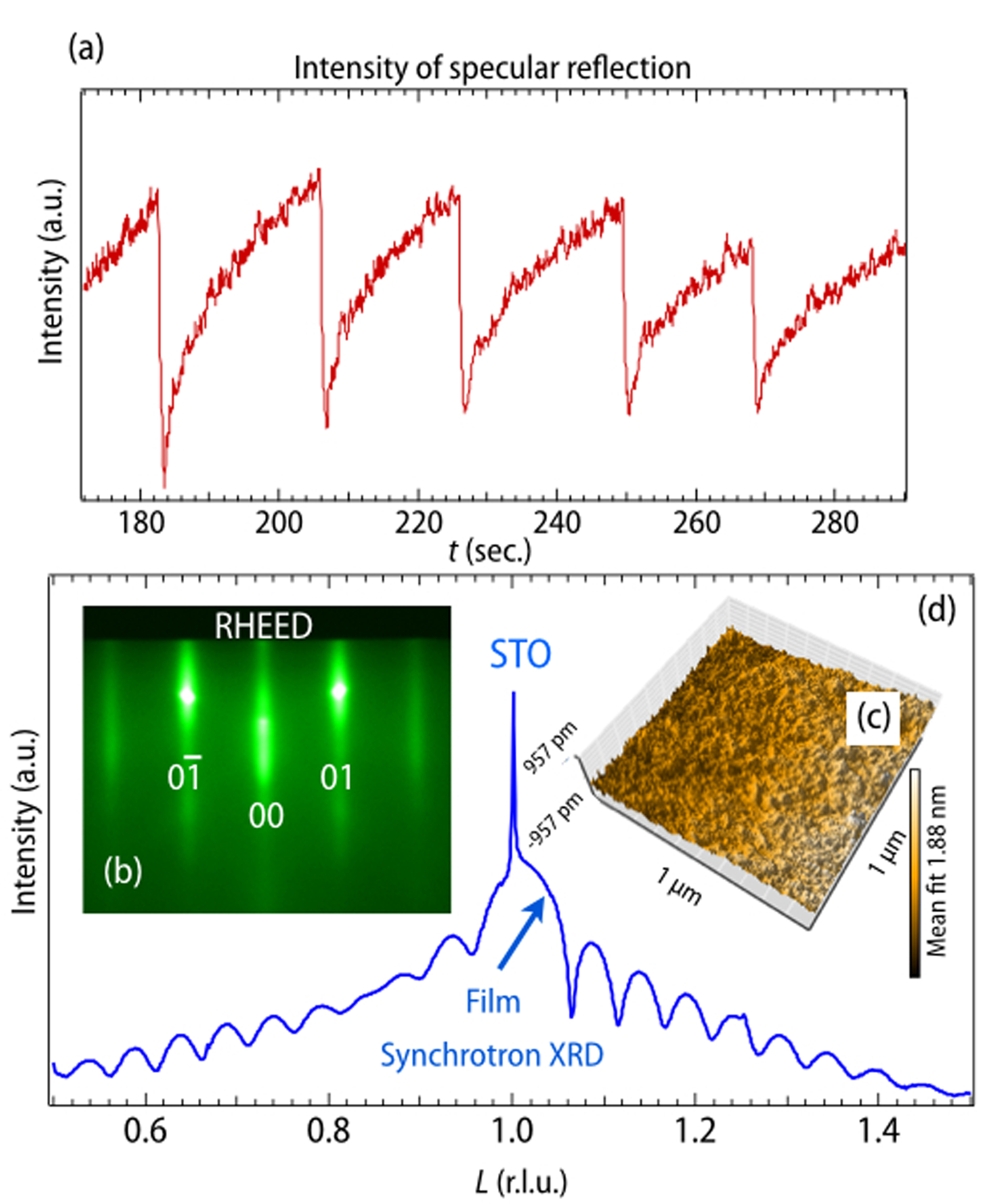}
\caption{\label{} (Color online) (a) RHEED specular intensity during the growth on STO substrate. (b) RHEED pattern of the 0th Laue circle on the same LCeMO film obtained after cooling to room temperature. The RHEED images were obtained along [010] direction. (c) AFM image (d) XRD pattern around STO (001) reflection.}
\end{figure}

  \begin{figure*}  
  \includegraphics[width=1\textwidth] {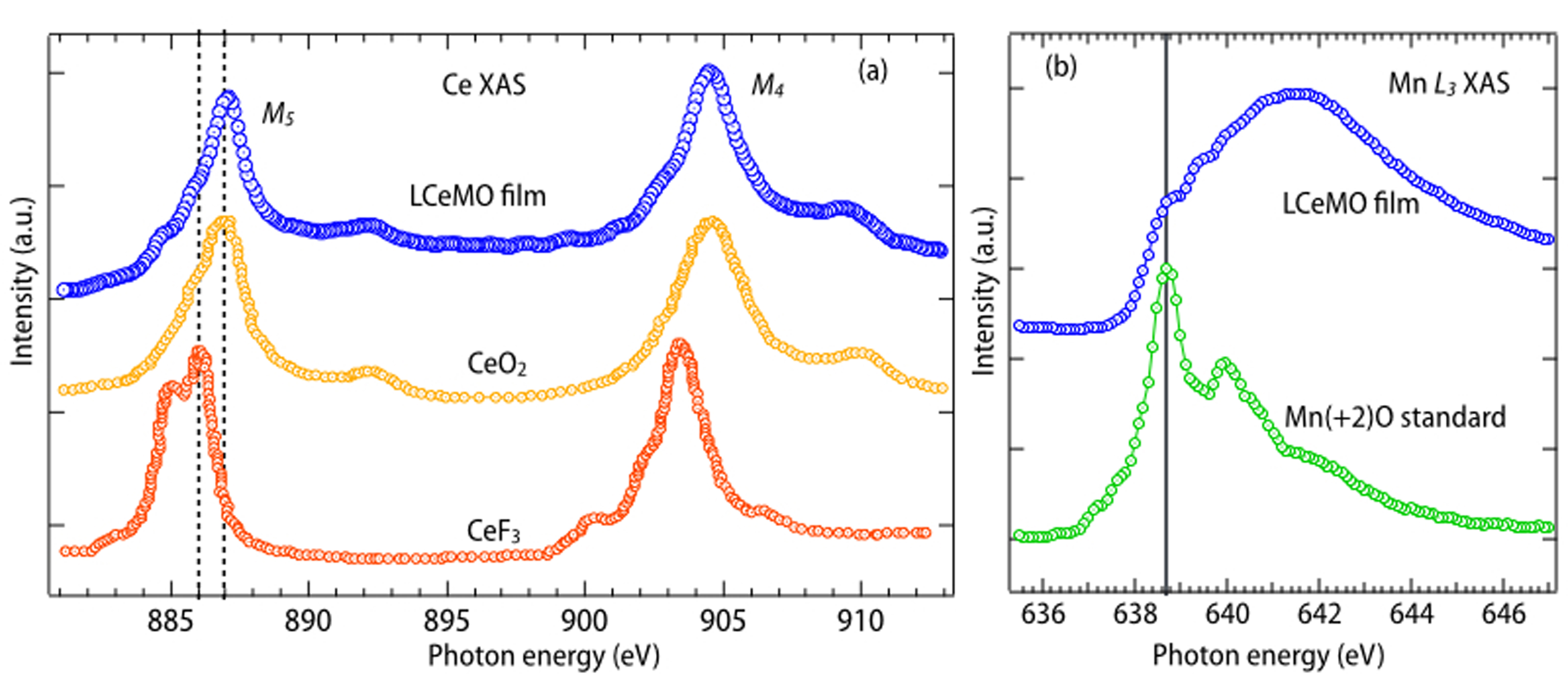}
\caption{\label{}  (Color online) (a) Ce $M_{5,4}$ XAS spectrum of La$_{0.8}$Ce$_{0.2}$MnO$_3$ film has been compared with the spectra of CeO$_2$ and CeF$_3$ standard, reproduced from Ref. ~\cite{lcemo_xas_edope}. (b) Mn $L_3$ XAS  spectra of La$_{0.8}$Ce$_{0.2}$MnO$_3$ film and Mn(+2)O standard.  All of the data in both panel (a) and (b) are moved arbitrarily along y axis for visual clarity. }
\end{figure*}

\begin{figure}  
\includegraphics[width=.47\textwidth] {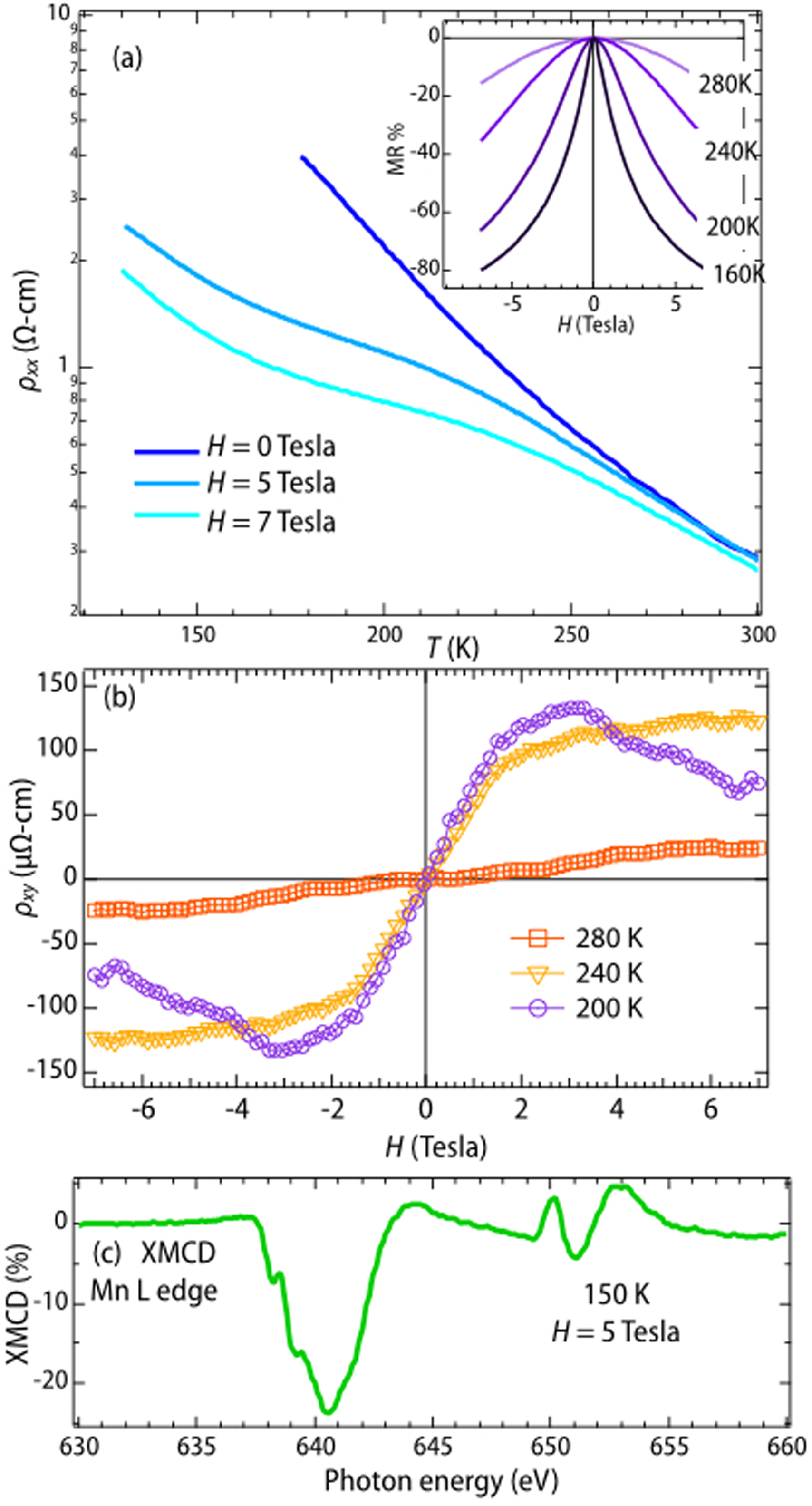}
\caption{\label{}  (Color online) (a) Resistivity vs. temperature for 20 uc LCeMO film measured with different magnetic field. The inset shows magnetoresistance (MR) as a function $H$ at several $T$. MR is defined as MR($H$) = $\rho_{xx}$  (b) Field dependence of the Hall resistivity measured at different $T$. (c)  XMCD
measurements at the Mn $L_{3,2}$ edges.}
\end{figure}



\begin{thebibliography}{99}
\bibitem{Hall} Here we recap,  that for a ferromagnet, the Hall resistivity~\cite{anomalous} is given by $\rho_{\textrm{xy}}$ = $\mu_0$ [$R_s M$ + $R_H H$] with $R_s$, $R_H$ being the anomalous and the ordinary Hall coefficients respectively,  where $\mu_0$ is the vacuum permeability and $M$ is the magnetization.  In order to remove the magnetoresistance contribution which are an even function of $H$, $\rho_{\textrm{xy}}$ has been obtained from the measured Hall resistivity ($\rho_{\textrm{raw}}$) as follows  $\rho_{\textrm{xy}}$= [$\rho_{\textrm{raw}}(\uparrow,+)$ - $\rho_{\textrm{raw}}(\downarrow,-)$ ]/2, where $\downarrow$, $\uparrow$, +, - present the increasing, decreasing, positive, negative external $H$, respectively~\cite{Hall1,Hall2}.
\end{thebibliography}
 \end{document}